\documentclass[prl,twocolumn,groupedaddress]{revtex4}
\usepackage{graphicx,color}
\usepackage{amssymb}   
\usepackage{amsmath}
\usepackage{epstopdf}
\usepackage{natbib}
\usepackage{hyperref}
\usepackage{bm}
\usepackage{color}
\usepackage{braket}

\begin{document}
\title{Spin-orbit coupling induced valley Hall effects in transition-metal dichalcogenides}
\author{Benjamin T. Zhou$^{1}$} \thanks{These authors contributed equally to this work.}
\author{Katsuhisa Taguchi$^{2}$} \thanks{These authors contributed equally to this work.}
\author{Yuki Kawaguchi$^{2}$, Yukio Tanaka$^{2}$}
\author{K. T. Law$^{1}$} \thanks{Corresponding author.\\phlaw@ust.hk}

\affiliation{$^{1}$Department of Physics, Hong Kong University of Science and Technology, Clear Water Bay, Hong Kong, China \\$^{2}$Department of Applied Physics, Nagoya University, Nagoya 464-8603, Japan }

\begin{abstract}
\bf{
In transition-metal dichalcogenides, electrons in the $K$-valleys can experience both Ising and Rashba spin-orbit couplings. In this work, we show that the coexistence of Ising and Rashba spin-orbit couplings leads to a special type of valley Hall effect, which we call spin-orbit coupling induced valley Hall effect. Importantly, near the conduction band edge, the valley-dependent Berry curvatures generated by spin-orbit couplings are highly tunable by external gates and dominate over the intrinsic Berry curvatures originating from orbital degrees of freedom under accessible experimental conditions. We show that the spin-orbit coupling induced valley Hall effect is manifested in the gate dependence of the valley Hall conductivity, which can be detected by Kerr effect experiments.}
\end{abstract}

\maketitle

\section*{Introduction}

Valley degrees of freedom emerge from local extrema in electronic band structures of two-dimensional Dirac materials. When spatial inversion symmetry is broken in such systems, valley-contrasting effective magnetic fields can arise in momentum space, known as Berry curvature fields\cite{Niu, Nagaosa}. Upon application of in-plane electric fields, the Berry curvature drives carriers from opposite valleys to flow in opposite transverse directions, leading to valley Hall effects (VHEs)\cite{Xiao, Tarucha}. It was first predicted that VHEs can exist in gapped graphene materials, where global inversion breaking is introduced by $h$-BN substrates\cite{Gorbachev} or external electric fields\cite{Sui, Shimazaki}. More recently, valley Hall phenomena were proposed in monolayer transition-metal dichalcogenides (TMDs)\cite{Xiao2}, in which nontrivial Berry curvatures result from intrinsically broken inversion symmetry in the trigonal prismatic structure of their unit cells\cite{Mattheiss}. Because of its versatility to couple to optical\cite{Heinz, Iwasa1, Fai1, Kim, Sie, Morpurgo, Onga, Cao}, magnetic\cite{Li, MacNiell, Aivazian, Srivastava, Tong} and electrical\cite{YeE, Fai2} controls, valley Hall physics in TMD-based materials have been under intensive theoretical and experimental studies in recent years.

Besides Berry curvature fields, broken inversion symmetry in monolayer TMDs also induces effective Zeeman fields in momentum space\cite{Xiao2, Yuan, Zeng, Zhu, Kormanyos, Zahid, Cappelluti, Kosmider, Liu}, referred to as Ising spin-orbit coupling(SOC) fields\cite{Ye1, Fai3, Zhou}. In the conduction band, the energy splitting due to Ising SOCs ranges from a few to tens of meVs\cite{Kosmider,Liu}, while in the valence bands it can be as large as $400$ meV in tungsten-based TMDs\cite{Yuan, Zeng}. Originating from in-plane mirror symmetry breaking and atomic spin-orbit interactions, the Ising SOC pins electron spins near opposite $K$-valleys to opposite out-of-plane directions. Due to its special roles in extending valley lifetimes\cite{Xiao2}, integrating spin and valley degrees of freedom\cite{Wang, Xie}, and enhancing upper critical fields in Ising superconductors\cite{Ye1, Fai3, Iwasa2, Houzet}, Ising SOCs have attracted extensive interests in studies of both valleytronics\cite{Heinz} and novel superconducting states\cite{Zhou, Noah, Wenyu, Sosenko, Eun-Ah, CXLiu} in TMD materials. In gated TMDs or polar TMDs\cite{Ang-Yu, Jing}, Rashba-type SOCs\cite{Rashba} also arise naturally\cite{Ye1, Roldan, Kormanyos2, Wan, Cheng}. Despite its wide existence, the effect of Rashba SOCs in TMDs has only been studied very lately, with focuses on superconducting states\cite{Ye1, Noah} and spintronic applications\cite{Klinovaja, Shao, Wan, Taguchi}.

In this work, we show that the coexistence of Ising and Rashba SOCs in gated/polar TMDs results in novel valley-contrasting Berry curvatures and a special type of valley Hall effect, which we call spin-orbit coupling induced valley Hall effects (SVHEs). In contrast to conventional Berry curvatures due to inversion-asymmetric hybridization of different $d$-orbitals\cite{Xiao2}, the new type of Berry curvatures originates from inversion-asymmetric spin-orbit interactions. To distinguish their physical origins, we refer to the Berry curvature induced by SOCs as spin-type Berry curvatures, and the conventional Berry curvatures/valley Hall effects from orbital degrees of freedom as orbital-type Berry curvatures/orbital VHEs. Importantly, under experimentally accessible gating\cite{Ye1, Iwasa2}, spin-type Berry curvatures near the conduction band edge can reach nearly ten times of its orbital counterpart. Thus, in gated or polar TMDs the SVHE can dominate over the orbital VHE, which significantly enhances the valley Hall effects in a wide class of TMDs and enriches the valley Hall phenomena in 2D Dirac materials. In addition, the SVHE proposed in this study provides a novel scheme to manipulate valley degrees of freedom of TMD materials. 

\section*{Results}
\subsection*{Massive Dirac Hamiltonian and spin-type Berry curvatures from Ising and Rashba SOCs}

\begin{figure}
\centering
\includegraphics[width=3.5in]{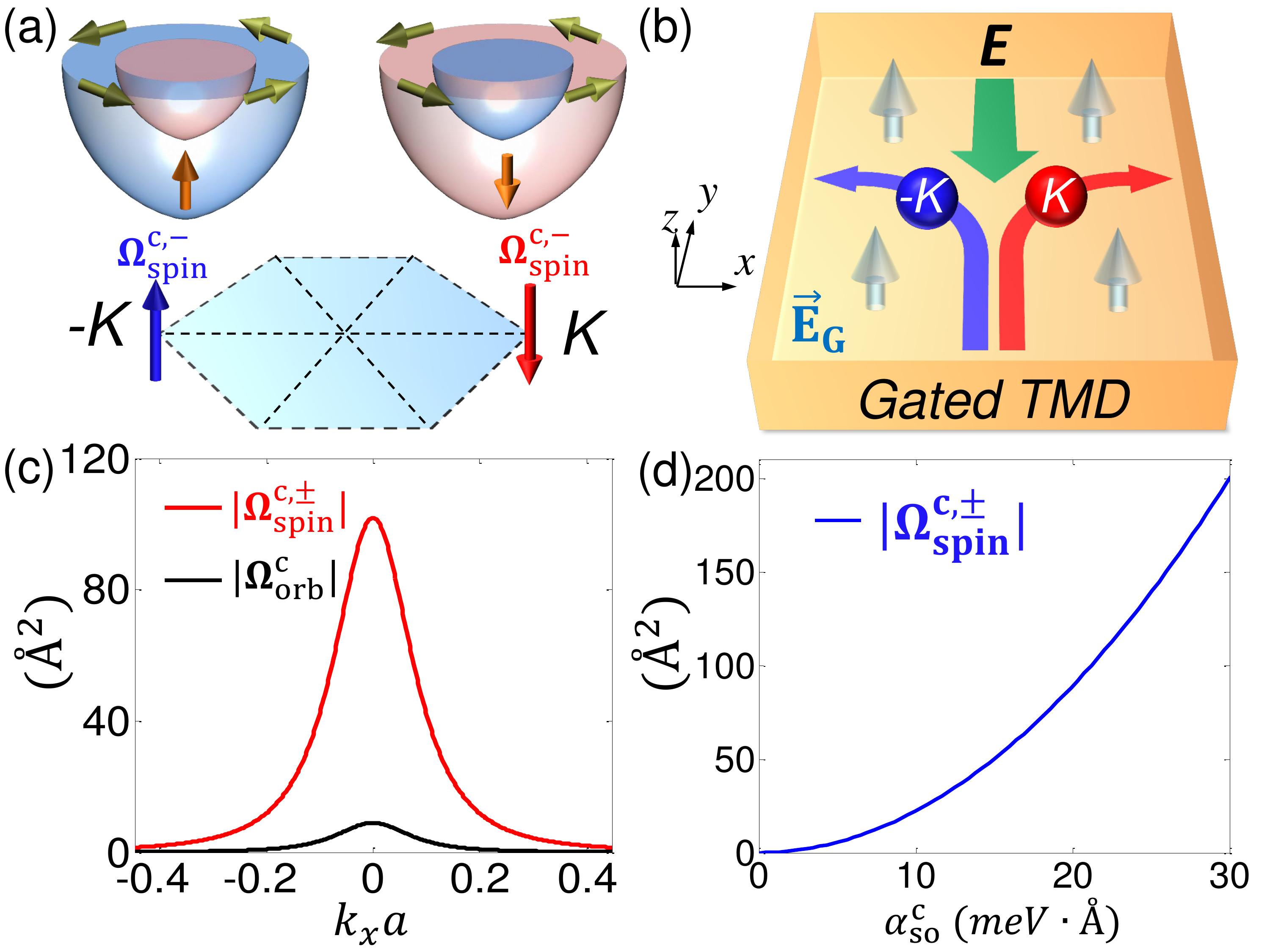}
\caption{Spin-orbit coupling induced valley Hall effects. (a) Schematics for the Ising spin-orbit coupling (SOC) (orange arrows), the Rashba SOC (golden arrows) and the spin-type Berry curvatures $\Omega^{\textrm{c},-}_{\textrm{spin}}$ (red/blue arrows) in the lower spin bands represented by red/blue pockets above $K$/$-K$-points.  (b) Valley Hall effects due to $\Omega^{\textrm{c}}_{\textrm{spin}}$. White arrows indicate out-of-plane gating fields/electric polarization labelled by $\textbf{E}_{\textrm{G}}$. (c) Magnitudes of spin-type Berry curvature $|\Omega^{\textrm{c},\pm}_{\textrm{spin}}|$ near the conduction band edge (red solid curve) and orbital-type Berry curvature $|\Omega^{\textrm{c}}_{\textrm{orb}}|$ (black solid curve) near $K$-points. Rashba coupling strength is set to be $\alpha^{\textrm{c}}_{\textrm{so}} = 21.4$ meV$\cdot{\AA}$ according to $\alpha^{\textrm{c}}_{\textrm{so}}k_{\textrm{F}}\approx 3$ meV\cite{Ye1}, comparable to $2|\beta^{\textrm{c}}_{\textrm{so}}| = 3$ meV\cite{Kosmider, Liu}. Parameters for $|\Omega^{\textrm{c}}_{\textrm{orb}}|$ are set to be: $\Delta = 0.83$ eV, $V_{\textrm{F}} = 3.51$ eV$\cdot{\AA}$\cite{Xiao2}. Clearly, $|\Omega^{\textrm{c},\pm}_{\textrm{spin}}|$ is nearly ten times of $|\Omega^{\textrm{c}}_{\textrm{orb}}|$. (d)$|\Omega^{\textrm{c},\pm}_{\textrm{spin}}|$ as a function of $\alpha^{\textrm{c}}_{\textrm{so}}$ at the $K$-points. Evidently, $|\Omega^{\textrm{c},\pm}_{\textrm{spin}}|$ scales quadratically with $\alpha^{\textrm{c}}_{\textrm{so}}$. }
\label{FIG01}
\end{figure}

To illustrate the spin-type Berry curvature and spin-orbit coupling induced valley Hall effect(SVHE) in TMDs, we consider gated monolayer MoS$_{2}$ as an example throughout this section, but the predicted effects generally exists in the whole class of gated TMDs or polar TMDs. In recent experiments, upon electrostatic gating the conduction band minima near the $K$-valleys can be partially filled\cite{Ye1, Iwasa2}, where the electron bands originate predominantly from the $4d_{z^2}$-orbitals of Mo-atoms\cite{Cappelluti, Liu}. Under the basis formed by spins of $d_{z^2}$-electrons, the effective Hamiltonian near the $K$-valleys for gated MoS$_{2}$ can be written as\cite{Ye1, Noah, Taguchi}:
\begin{eqnarray}\label{eq:01}
H_{\textrm{spin}}(\bm{k} + \epsilon \bm{K}) = \xi^{\textrm{c}}_{\bm{k}} \sigma_0 + \alpha^{\textrm{c}}_{\textrm{so}} (k_y \sigma_x - k_x \sigma_y) + \epsilon \beta^{\textrm{c}}_{\textrm{so}} \sigma_z .
\end{eqnarray}

Here, $\xi^{\textrm{c}}_{\bm{k}} =\frac{|\bm{k}|^2}{2m^{\ast}_{\textrm{c}}} - \mu$ denotes the usual kinetic energy term, $m^{\ast}_{\textrm{c}}$ is the effective mass of the electron band, $\mu$ is the chemical potential, $\bm{k}=(k_x, k_y)$ is the momentum displaced from $K (-K)$-valleys, $\epsilon=\pm$ is the valley index. The $\beta^{\textrm{c}}_{\textrm{so}}$-term refers to the Ising SOC which pins electron spins to out-of-plane directions (depicted by the orange arrows in Fig.\ref{FIG01}a). The origin of Ising SOC is the breaking of an in-plane mirror symmetry (mirror plane perpendicular to the 2D lattice plane) as well as the atomic SOC from the transition metal atoms. The $\alpha^{\textrm{c}}_{\textrm{so}}$-term describes the Rashba SOC which pins electron spins in in-plane directions with helical spin textures (indicated by the golden arrows in Fig.\ref{FIG01}a). Rashba SOC will arise when the out-of-plane mirror symmetry (mirror plane parallel to the lattice plane) is broken by gating or by lattice structure (as in the case of polar TMDs \cite{Ang-Yu, Jing}). Clearly, $H_{\textrm{spin}}$ has the form of a massive Dirac Hamiltonian (by neglecting the kinetic term which has no contribution to Berry curvatures), and the Ising SOC plays the role of a valley-contrasting Dirac mass, which is on the order of a few to tens of meVs\cite{Liu}.  

Importantly, the Pauli matrices $\bm{\sigma} = (\sigma_x, \sigma_y, \sigma_z)$ in Eq.\ref{eq:01} act on spin degrees of freedom. This stands in contrast to the massive Dirac Hamiltonian in Refs.\cite{Xiao2}:
\begin{eqnarray}\label{eq:Horbital}
H_{\textrm{orb}}(\bm{k} + \epsilon \bm{K}) = V_{\textrm{F}} ( \epsilon k_x \tau_x + k_y \tau_y) + \Delta \tau_z .
\end{eqnarray}
where the Pauli matrices $\bm{\tau} = (\tau_x, \tau_y, \tau_z)$ act on the subspace formed by different $d$-orbitals. The $V_{\textrm{F}}$-term results from electron hopping, and the Dirac mass $\Delta$ is generated by the large band gap ($\sim 2\Delta$) on the order of $1-2 eV$s in monolayer TMDs\cite{Xiao2}.

As shown in Fig.\ref{FIG01}a, Ising and Rashba SOCs result in non-degenerate spin sub-bands near the conduction band minimum. The energy spectra of upper/lower spin-subbands are given by $E^{\epsilon}_{\textrm{c},\pm}(\bm{k})=\xi^{\textrm{c}}_{\bm{k}} \pm \sqrt{(\alpha^{\textrm{c}}_{\textrm{so}}k)^2 + (\epsilon\beta^{\textrm{c}}_{\textrm{so}})^2}$. The Berry curvatures generated by SOCs in the lower spin-bands with energy $E^{\epsilon}_{c,-}(\bm{k})$ is given by:
\begin{eqnarray}\label{eq:02}
\Omega^{\textrm{c},-}_{\textrm{spin}}(\bm{k}+\epsilon \bm{K}) =  \frac{ (\alpha^{\textrm{c}}_{\textrm{so}})^2 \epsilon\beta^{\textrm{c}}_{\textrm{so}}}{2 [(\alpha^{\textrm{c}}_{\textrm{so}}k)^2 + (\beta^{\textrm{c}}_{\textrm{so}})^2]^{3/2}} .
\end{eqnarray}

Note that $\Omega^{\textrm{c},-}_{\textrm{spin}}$ has valley-dependent signs due to the valley-contrasting Dirac mass generated by Ising SOCs. As a result, under an in-plane electric field, $\Omega^{\textrm{c},-}_{\textrm{spin}}$ can drive electrons in the lower spin-bands at opposite valleys to flow in opposite transverse directions, which leads to transverse valley currents(Fig.\ref{FIG01}b). To distinguish this novel phenomenon from the intrinsic VHE in monolayer TMDs\cite{Xiao2}, we call this special type of VHE the spin-orbit coupling induced valley Hall effect (SVHE) due to its physical origin in spin degrees of freedom. Likewise, the Berry curvatures generated by Ising and Rashba SOCs are called spin-type Berry curvatures to distinguish it from the orbital-type Berry curvatures due to inversion-asymmetric mixing of different $d$-orbitals\cite{Xiao2}. 

We note that for the upper spin-band with energy $E^{\epsilon}_{\textrm{c},+}(\bm{k})$, we have $\Omega^{\textrm{c},+}_{\textrm{spin}} = -\Omega^{\textrm{c},-}_{\textrm{spin}}$. Therefore, valley currents from upper and lower spin-bands can partially cancel each other when both of them are occupied. However, non-zero valley currents can still be generated due to the population difference in the spin-split bands.

Based on Eq.\ref{eq:02}, $\Omega^{\textrm{c},\pm}_{\textrm{spin}}$ has a formal similarity with its orbital counterpart\cite{Xiao2}:
\begin{eqnarray}\label{eq:OmegaOrbital}
\Omega^{\textrm{c}}_{\textrm{orb}}(\bm{k}+\epsilon \bm{K}) =  -\frac{\epsilon V_{\textrm{F}}^2 \Delta}{2 [(V_{\textrm{F}} k)^2 + \Delta^2]^{3/2}} .
\end{eqnarray}
However, we point out that $\Omega^{\textrm{c},\pm}_{\textrm{spin}}$ originates from a very different physical mechanism from $\Omega^{\textrm{c}}_{\textrm{orb}}$ and has important implications in valleytronic applications. 

On one hand, the magnitude of $\Omega^{\textrm{c}}_{\textrm{orb}}$ in TMDs is generally small ($\sim 10$ ${\AA^2}$) due to the large Dirac mass from the band gap $2\Delta\sim 1-2$ eV\cite{Xiao2}. In contrast, for $\Omega^{\textrm{c},\pm}_{\textrm{spin}}$, the Dirac mass $\beta^{\textrm{c}}_{\textrm{so}}$ is on the order of a few meVs near the conduction band edges\cite{Liu}. For gated MoS$_2$, the Rashba energy can reach $\alpha^{\textrm{c}}_{\textrm{so}}k_{\textrm{F}} \approx 3$ meV at the Fermi energy\cite{Ye1} (see Supplementary Note 1 for details), which is comparable to the energy-splitting $2|\beta^{\textrm{c}}_{\textrm{so}}|\approx 3$ meV caused by Ising SOCs\cite{Liu}. In this case, $|\Omega^{\textrm{c},\pm}_{\textrm{spin}}|$ near the conduction band minimum can be nearly ten times of $|\Omega^{\textrm{c}}_{\textrm{orb}}|$ (Fig.\ref{FIG01}c). Therefore, the SVHE is expected to generate pronounced valley Hall signals in gated/polar TMDs.

On the other hand, the strength of $\Omega^{\textrm{c}}_{\textrm{orb}}$ is determined by parameters intrinsic to the material, thus can hardly be tuned. However, $\Omega^{\textrm{c},\pm}_{\textrm{spin}}$ has a quadratic dependence on the Rashba coupling strength $\alpha^{\textrm{c}}_{\textrm{so}}$ (Eq.\ref{eq:02}), which can be controlled by external gating fields. As shown in Fig.\ref{FIG01}d, $|\Omega^{\textrm{c},\pm}_{\textrm{spin}}|$ can be strongly enhanced by increasing $\alpha^{\textrm{c}}_{\textrm{so}}$ within experimentally accessible gating strength\cite{Ye1}. This suggests that the SVHE can serve as a promising scheme for electrical control of valleys in TMD-based valleytronic devices. 

We note that the form of effective Hamiltonian in Eq.\ref{eq:01} also applies to the $K$-valleys in the valence band (see Supplementary Note 2), thus SVHEs can also occur in the valence band. Unfortunately, as we demonstrate below, the spin-type Berry curvature is much weaker in the valence band due to the giant Ising SOC strength $\beta^{\textrm{v}}_{\textrm{so}}\sim 100$ meV near the valence band edge\cite{Xiao2, Yuan, Zeng, Zhu, Kormanyos, Zahid, Cappelluti, Liu}.
 
\begin{figure}
\centering
\includegraphics[width=3.5in]{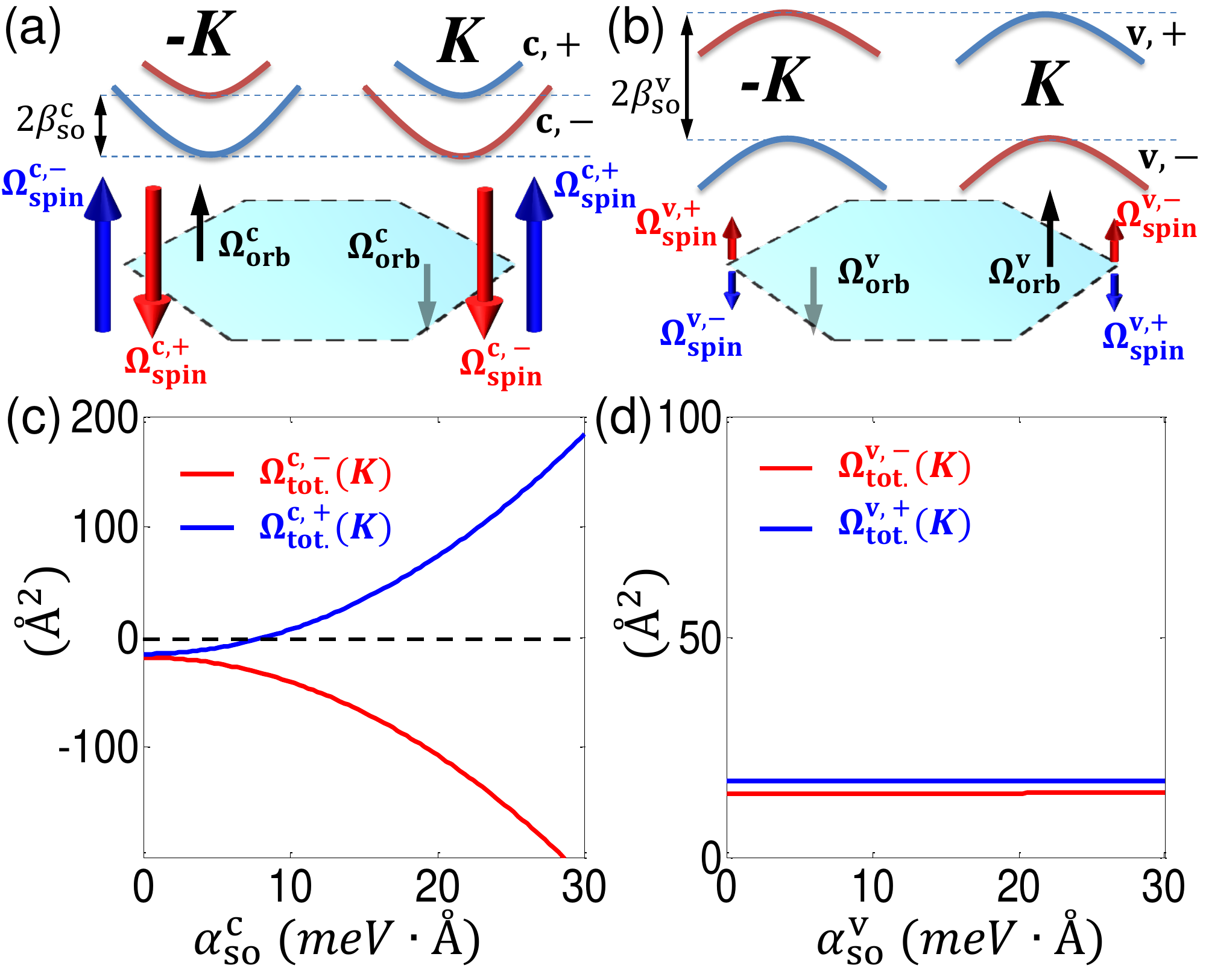}
\caption{Interplay between $\Omega_{\textrm{spin}}$ and $\Omega_{\textrm{orb}}$. (a) The case near the conduction band edge. (b) The case near the valence band edge. (c) Total Berry curvatures $\Omega^{\textrm{c},\pm}_{\textrm{tot}.}$ of upper($\textrm{c}, +$)/lower($\textrm{c}, -$) spin-subbands at $+\bm{K}$-point versus $\alpha^{\textrm{c}}_{\textrm{so}}$ near the conduction band edge. As $\alpha^{\textrm{c}}_{\textrm{so}}$ increases, $\Omega^{\textrm{c},\pm}_{\textrm{spin}}$ becomes dominant and changes $\Omega^{\textrm{c},\pm}_{\textrm{tot}.}$ dramatically. (d) Total Berry curvatures $\Omega^{\textrm{v},\pm}_{\textrm{tot}.}$ of upper($\textrm{v}, +$)/lower($\textrm{v}, -$) spin-subbands at $+\bm{K}$-point as a function of $\alpha^{\textrm{v}}_{\textrm{so}}$ at the valence band edge. Obviously, $\Omega^{\textrm{v},\pm}_{\textrm{tot}.}$ are insensitive to $\alpha^{\textrm{v}}_{\textrm{so}}$ and remain close to $\Omega^{\textrm{v}, \pm}_{\textrm{orb}}$ at $\alpha^{\textrm{v}}_{\textrm{so}} = 0$.}
\label{FIG02}
\end{figure}

\subsection*{Interplay between spin-type and orbital-type Berry curvatures}

In real gated/polar TMDs, the spin-type Berry curvature $\Omega_{\textrm{spin}}$ always coexist with the orbital-type Berry curvature $\Omega_{\textrm{orb}}$. In this section, we demonstrate the interplay between $\Omega_{\textrm{spin}}$ and $\Omega_{\textrm{orb}}$ near the $K$-valleys (shown schematically in Fig.\ref{FIG02}a-b). Specifically, using monolayer MoS$_2$ as an example, we study the total Berry curvatures at the $K$-points based on a realistic tight-binding(TB) model\cite{Liu} which takes both $\Omega_{\textrm{spin}}$ and $\Omega_{\textrm{orb}}$ into account. The TB Hamiltonian is presented in the Method section and detailed model parameters are presented in the Supplementary Note 3.

For simplicity, we focus on Berry curvatures at $\bm{K}=(4\pi/3a,0)$, and the physics at $-\bm{K}$ follows naturally from the requirement imposed by time-reversal symmetry: $\Omega(-\bm{K}) = -\Omega(\bm{K})$. 

First, we study the conduction band case where the Ising SOC strength $\beta^{\textrm{c}}_{\textrm{so}}$ is small. In the absence of gating, the total Berry curvatures $\Omega^{\textrm{c},\pm}_{\textrm{tot}.}$ in both spin-subbands at $\bm{K}=(4\pi/3a,0)$ consist of orbital-type contributions $\Omega^{\textrm{c},\pm}_{\textrm{orb}}$ only, both pointing to the negative $z$-direction\cite{Xiao2}. By gradually turning on the Rashba coupling strength, $\Omega^{\textrm{c},\pm}_{\textrm{spin}}$ come into play and change $\Omega^{\textrm{c},\pm}_{\textrm{tot}.}$ dramatically. In particular, for the lower spin-subband, $\Omega^{\textrm{c},-}_{\textrm{spin}}$ also points to the negative $z$-direction (Fig.\ref{FIG02}a). This is due to the fact that $\beta^{\textrm{c}}_{\textrm{so}}<0$ in molybdenum(Mo)-based TMDs\cite{Kosmider, Liu}, which leads to a negative value of $\Omega^{\textrm{c},-}_{\textrm{spin}}$ (Eq.\ref{eq:02}). As a result, $\Omega^{\textrm{c},-}_{\textrm{tot}.}$ keeps increasing its magnitude as $\alpha^{\textrm{c}}_{\textrm{so}}$ increases (red solid curve in Fig.\ref{FIG02}c). For the upper spin-subband, however, $\Omega^{\textrm{c},+}_{\textrm{spin}}=-\Omega^{\textrm{c},-}_{\textrm{spin}}$, which is anti-parallel to its orbital counterpart $\Omega^{\textrm{c},+}_{\textrm{orb}}$, thus they compete against each other as shown in Fig.\ref{FIG02}a. As $\alpha^{\textrm{c}}_{\textrm{so}}$ grows up, $\Omega^{\textrm{c},+}_{\textrm{spin}}$ becomes comparable to $\Omega^{\textrm{c},+}_{\textrm{orb}}$, resulting in a zero total Berry curvature $\Omega^{\textrm{c},+}_{\textrm{tot}.} = 0$ at certain $\alpha^{\textrm{c}}_{\textrm{so}}$ (indicated by the intersection between the blue curve and zero in Fig.\ref{FIG02}c). As $\alpha^{\textrm{c}}_{\textrm{so}}$ increases further, $\Omega^{\textrm{c},\pm}_{\textrm{spin}}$ dominates, leading to different signs of total Berry curvatures in upper and lower spin bands, with $\Omega^{\textrm{c},+}_{\textrm{tot}.}>0$ and $ \Omega^{\textrm{c},-}_{\textrm{tot}.}<0$.

Notably, in tungsten(W)-based TMDs Ising SOCs in the conduction band have a different sign $\beta^{\textrm{c}}_{\textrm{so}}>0$\cite{Kosmider, Liu}. In this case, $\Omega^{\textrm{c},-}_{\textrm{spin}}$ competes with $\Omega^{\textrm{c},-}_{\textrm{orb}}$, while $\Omega^{\textrm{c},+}_{\textrm{spin}}$ aligns with $\Omega^{\textrm{c},+}_{\textrm{orb}}$. This is contrary to the behaviors in molybdenum(Mo)-based materials. As a result, $\Omega^{\textrm{c},-}_{\textrm{tot}.}$ in W-based materials can change its sign when $\Omega^{\textrm{c},-}_{\textrm{spin}}$ dominates, similar to $\Omega^{\textrm{c},+}_{\textrm{tot}.}$ in Mo-based case (blue curve in Fig.\ref{FIG02}c). As we discuss in the next section, this can reverse the direction of total valley currents. Similar plots as shown in Fig.\ref{FIG02} for W-based TMDs are presented in Supplementary Note 4.

In contrast to conduction bands, valence band edges in TMDs exhibit extremely strong Ising SOC  with $\beta^{\textrm{v}}_{\textrm{so}}\sim 100 - 200$ meV\cite{Xiao2, Yuan, Zeng, Zhu, Kormanyos, Zahid, Cappelluti}. This leads to very weak spin-type Berry curvatures $\Omega^{\textrm{v},\pm}_{\textrm{spin}}$ (shown schematically in Fig.\ref{FIG02}b). This is because electron spins near the valence band edges are strongly pinned by the Ising SOCs to the out-of-plane directions, and the Rashba SOCs due to gating or electric polarization cannot compete with Ising SOCs. As a result, the Berry phase acquired during an adiabatic spin rotation driven by Rashb SOC fields becomes negligible. Therefore, in the valence band the orbital-type contribution $\Omega^{\textrm{v},\pm}_{\textrm{orb}}$ generally dominates. As shown clearly in Fig.\ref{FIG02}d, $\Omega^{\textrm{v},\pm}_{\textrm{tot}.}$ are almost insensitive to $\alpha^{\textrm{v}}_{\textrm{so}}$ and remain close to $\Omega^{\textrm{v},\pm}_{\textrm{orb}}$ at $\alpha^{\textrm{v}}_{\textrm{so}}=0$.

It is worth noting that the behavior of total Berry curvatures in Fig.\ref{FIG02}(c)-(d) can be understood by considering spin-type and orbital-type contributions separately. This is due to the fact that the total Berry curvature $\Omega^{n}_{\textrm{tot}.}$ at the $K$-points for a given band $n$ can be written as the algebraic sum of $\Omega^{n}_{\textrm{spin}}$ and $\Omega^{n}_{\textrm{orb}}$: $\Omega^{n}_{\textrm{tot}.}(\epsilon K) = \Omega^{n}_{\textrm{spin}}(\epsilon K) + \Omega^{n}_{\textrm{orb}}(\epsilon K)$. Detailed derivations can be found in Supplementary Note 5.

\subsection*{Detecting spin-orbit coupling induced valley Hall effects}

In this section, we discuss how to detect unique experimental signatures of SVHEs in $n$-type monolayer TMDs using Kerr effect measurements. In particular, we study the cases of molybdenum(Mo)-based and tungsten(W)-based TMDs separately. 

As demonstrated in the previous section, for Mo-based materials the total Berry curvature in the lower spin-band $\Omega^{\textrm{c},-}_{\textrm{tot}.}$ can be significantly enhanced by $\Omega^{\textrm{c},-}_{\textrm{spin}}$ under gating(Fig.\ref{FIG02}c). Therefore, when only the lower spin-bands are filled, the extra contribution from SVHEs can strongly enhance the total valley Hall conductivity $\sigma^{\textrm{V}}_{xy}$ in gated/polar TMDs, which is expected to far exceed the intrinsic $\sigma^{\textrm{V}}_{xy}$ from orbital VHEs. 

Moreover, when $\Omega^{\textrm{c}}_{\textrm{spin}}$ dominates, $\Omega^{\textrm{c},-}_{\textrm{tot}.}$ has a different sign (Fig.\ref{FIG02}c). When both spin-bands are filled, valley currents from upper and lower spin bands partially cancel each other, with finite valley currents generated from their population difference. In this case $\sigma^{\textrm{V}}_{xy}$ is expected to increase at a lower rate as doping level increases. This behavior is very different from the orbital valley Hall effect for electron-doped samples: since $\Omega^{\textrm{c},\pm}_{\textrm{orb}}$ have the same sign\cite{Xiao2}(See Fig.\ref{FIG02}(c) at $\alpha^{\textrm{c}}_{\textrm{so}}=0$), when both spin sub-bands are filled, $\sigma^{\textrm{V}}_{xy}$ is expected to increase at a higher rate as a function of doping level.

\begin{figure}
\centering
\includegraphics[width=3.5in]{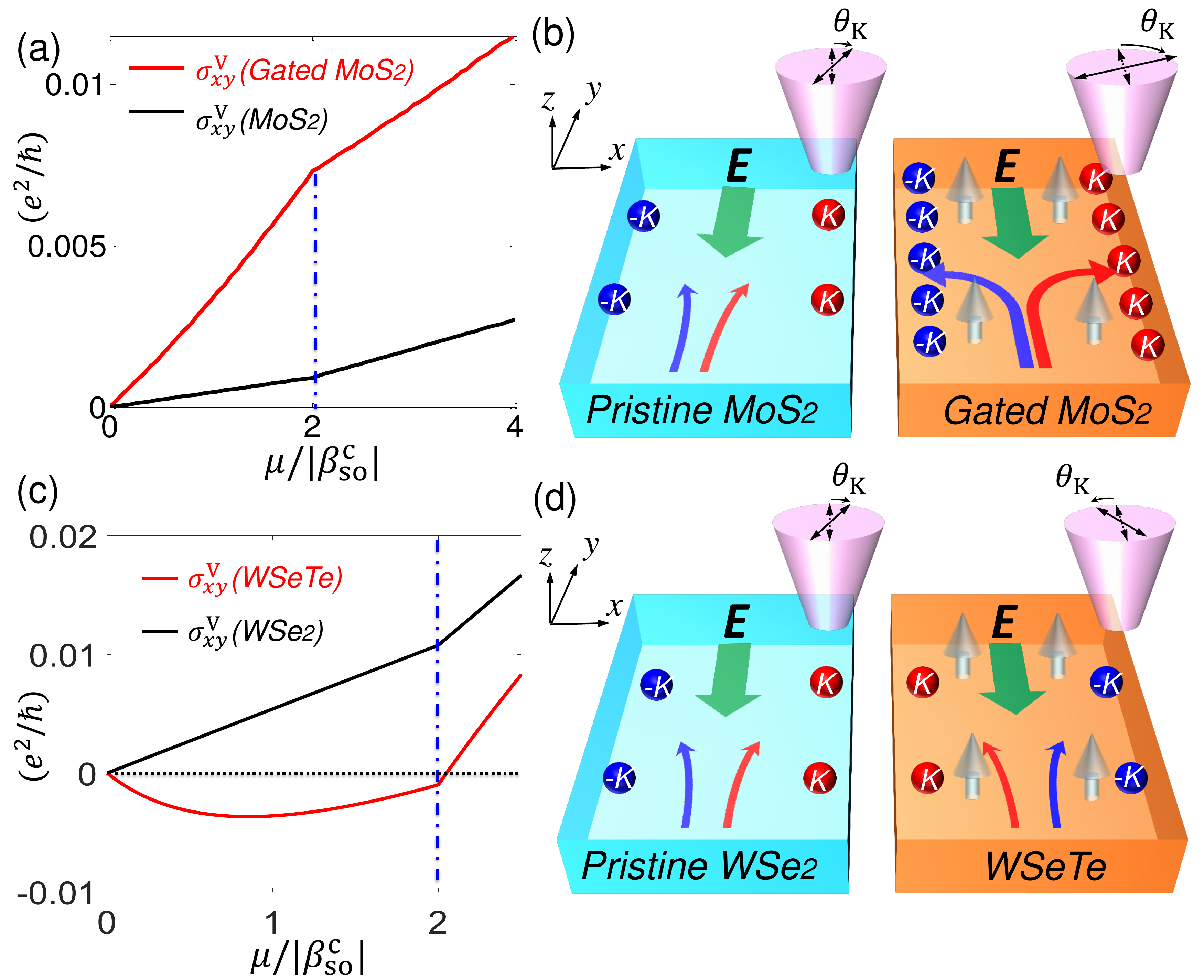}
\caption{Detecting spin-orbit coupling induced valley Hall effects(SVHEs). (a) Total valley Hall conductivity $\sigma^{V}_{xy}$ as a function of chemical potential $\mu$ for gated MoS$_2$ (red curve) and pristine sample (black curve). The blue dashed line indicates the location where $\mu = 2|\beta^{\textrm{c}}_{\textrm{so}}|$. (b) Polar Kerr effect measurements to detect SVHEs in Mo-based transition-metal dichalcogenides(TMDs). For Mo-based TMDs, SVHEs strongly enhance $\sigma^{\textrm{V}}_{xy}$ in the regime $\mu \sim 2|\beta^{\textrm{c}}_{\textrm{so}}|$ comparing to the intrinsic value. This creates a significant valley imbalance $n_{\textrm{V}}$ and valley magnetization at the sample boundaries, which can be signified by a large Kerr angle $\theta_{\textrm{K}}$. (c) Total $\sigma^{\textrm{V}}_{xy}$ versus chemical potential $\mu$ for polar TMD WSeTe (red curve) and pristine WSe$_2$ (black curve). Clearly, in the regime $\mu < 2|\beta^{\textrm{c}}_{\textrm{so}}|$ the sign of $\sigma^{\textrm{V}}_{xy}$ in WSeTe is reversed due to SVHEs. (d) Schematics for polar Kerr experiments to detect SVHEs in tungsten-based polar TMD WSeTe. The reversed valley current is signified by the sign reversal of $\theta_{\textrm{K}}$.}
\label{FIG03}
\end{figure}

To study this unique signature of $\sigma^{\textrm{V}}_{xy}$ due to SVHEs, we calculate $\sigma^{\textrm{V}}_{xy}$ for $n$-type monolayer MoS$_{2}$ using the tight-binding model\cite{Liu} presented in the Method section. The $\sigma^{\textrm{V}}_{xy}$ for electron-doped TMDs is given by (see Supplementary Note 6 for details):
\begin{eqnarray}\label{eq:03}
\sigma^{\textrm{V}}_{xy} = -\frac{2e^2}{\hbar} \int  \frac{d^2\bm{k}}{(2\pi)^2} [f_{\textrm{c},+}(\bm{k})\Omega^{\textrm{c},+}_{\textrm{tot}.}(\bm{k})+f_{\textrm{c},-}(\bm{k})\Omega^{\textrm{c},-}_{\textrm{tot}.}(\bm{k})].
\end{eqnarray}

Here, the integral is calculated near the $\bm{K}$-point, and $f_{\textrm{c},\pm}(\bm{k})=\{1+\exp[(E^{\epsilon=+}_{\textrm{c},\pm}(\bm{k})-\mu)/k_{\textrm{B}} T]\}^{-1}$ are the Fermi functions associated with the upper/lower spin-bands near the conduction band edge. In the limit $T \rightarrow 0$, the calculated $\sigma^{\textrm{V}}_{xy}$ as a function of chemical potential $\mu$ for gated (red solid curve) and pristine (black solid curve) monolayer MoS$_{2}$ are shown in Fig.\ref{FIG03}a. The chemical potential $\mu$ is measured from the conduction band minimum.

When $\mu < 2|\beta^{\textrm{c}}_{\textrm{so}}|$, only the lower spin-band is occupied, \textit{i.e.}, $f_{\textrm{c},+}(\bm{k})=0$. It is evident from Fig.\ref{FIG03}a that as $\mu$ increases, the net $\sigma^{\textrm{V}}_{xy}$ for gated MoS$_2$ (red solid curve in Fig.\ref{FIG03}a) grows much more rapidly than the intrinsic $\sigma^{\textrm{V}}_{xy}$ (black solid curve in Fig.\ref{FIG03}a). For $\mu>2|\beta^{\textrm{c}}_{\textrm{so}}|$, the intrinsic $\sigma^{\textrm{V}}_{xy}$ starts increasing at a slightly higher rate, while the $\sigma^{\textrm{V}}_{xy}$ for gated sample increases at a lower rate. 

To detect this distinctive flattening behavior in the $\sigma^{\textrm{V}}_{xy}-\mu$ curve due to SVHEs, we propose polar Kerr effect experiments (Fig.\ref{FIG03}b) which can directly map out the spatial profile of net magnetization in a 2D system\cite{Fai2, Lee}. In the steady state, valley currents $\textbf{J}_{\textrm{v}} = \sigma^{\textrm{V}}_{xy} \textbf{E} \times \hat{z}$ generated by the electric field $\textbf{E}$ (green arrows in Fig.\ref{FIG03}b) are balanced by valley relaxations at the sample boundaries, which establishes a finite valley imbalance $n_{\textrm{V}} \propto \sigma^{\textrm{V}}_{xy}$ near the sample edges. Due to valley-contrasting Berry curvatures, the valley imbalance $n_{\textrm{V}}$ induces a nonzero out-of-plane orbital magnetization $M_{z} \propto n_{\textrm{V}}$\cite{Niu}, which can be measured by the Kerr rotation angle $\theta_{\textrm{K}}$, with $\theta_{\textrm{K}} \propto n_{\textrm{V}} \propto \sigma^{\textrm{V}}_{xy}$\cite{Fai2}. Therefore, $\theta_{\textrm{K}}$ as a function of doping level for intrinsic/gated monolayer TMDs are expected to exhibit similar features as the $\sigma^{\textrm{V}}_{xy} - \mu$ curves in Fig.\ref{FIG03}a. In previous Kerr measurements on MoS$_2$, the Kerr angle due to valley imbalance generated by orbital VHE is roughly $\theta_{\textrm{K}}\approx 60$ $\mu$rad when both subbands are filled\cite{Fai2}. According to Fig.\ref{FIG03}a, the SVHE can enhance $\sigma^{\textrm{V}}_{xy}$ by nearly 4-5 times when $\mu>2|\beta^{\textrm{c}}_{\textrm{so}}|$, hence we expect $\theta_{\textrm{K}} \approx 200-300$ $\mu$rad at the same doping level for gated MoS$_2$. 

Now we discuss the distinctive signature of SVHE in $n$-type tungsten(W)-based TMDs. As we pointed out in the last section, $\Omega^{\textrm{c},-}_{\textrm{spin}}$ competes with $\Omega^{\textrm{c},-}_{\textrm{orb}}$ in W-based materials due to $\beta^{\textrm{c}}_{\textrm{so}} > 0$. Therefore, when $\Omega^{\textrm{c},-}_{\textrm{spin}}$ dominates, $\Omega^{\textrm{c},-}_{\textrm{tot}.}$ has an opposite sign to $\Omega^{\textrm{c},-}_{\textrm{orb}}$. When only the lower spin-band is filled (\textit{i.e.}, $\mu<2|\beta^{\textrm{c}}_{\textrm{so}}|$ and $f_{\textrm{c},+}(\bm{k})=0$), this changes the sign of $\sigma^{\textrm{V}}_{xy}$(Eq.\ref{eq:03}), and the direction of valley currents is reversed. 

To demonstrate the reversal of valley current directions due to SVHEs in W-based TMDs, we compare the $\sigma^{\textrm{V}}_{xy}$ of a pristine monolayer tungsten-diselenide(WSe$_2$) and a polar TMD tungsten-selenide-telluride(WSeTe) where strong band-splitting induced by Rashba SOCs is predicted\cite{Wan}. Using fitted values of $\alpha^{\textrm{c}}_{\textrm{so}}$ and $\beta^{\textrm{c}}_{\textrm{so}}$ for WSeTe (details presented in Supplementary Note 4), we calculate the $\sigma^{\textrm{V}}_{xy}-\mu$ curves for WSe$_2$ and WSeTe as shown in Fig.\ref{FIG03}c. Clearly, for $\mu<2|\beta^{\textrm{c}}_{\textrm{so}}|$, $\sigma^{\textrm{V}}_{xy}$ of WSeTe has a different sign from that of WSe$_2$. As a result, the valley currents in WSe$_2$ and WSeTe under applied electric field flow in opposite transverse directions(Fig.\ref{FIG03}d). This leads to opposite valley magnetization on the same boundaries, which can be signified by the sign difference in $\theta_{\textrm{K}}$ correspondingly\cite{Fai2}.

We note that the relation $\theta_{\textrm{K}} \propto n_{\textrm{V}}$ discussed above relies on the fact that the valley orbital magnetic moments remain almost unaffected by $\Omega_{\textrm{spin}}$. This is explained in details in Supplementary Note 7.

\section*{Discussion}

We discuss a few important aspects on SVHEs studied above. First, the novel SVHE as well as its unique signatures studied above applies in general to the whole class of monolayer TMDs. In particular, for molybdenum-based materials, strong Ising and Rashba SOC effects in the conduction band, such as MoSe$_2$ and MoTe$_2$\cite{Wan, Liu}, have sizable band-splitting of $20 - 30$ meV near the band edge and exhibit pronounced signals of SVHEs. Detailed calculations of SVHEs in MoTe$_2$ are presented in Supplmentary Note 8. 

Second, a strong gating field is not necessarily required to induce strong Rashba SOCs in TMDs. As we mentioned above, in polar transition-metal dichalcogenides MXY (M$=$Mo,W; X,Y$=$S, Se, Te and X $\neq$ Y)\cite{Wan, Cheng, Ang-Yu, Jing}, out-of-plane electric polarizations are bulit-in from intrinsic mirror symmetry breaking in the crystal structure. Thus, Ising and Rashba SOCs naturally coexist in these polar TMD materials without any further experimental design. This is very different from graphene-based devices where valley currents are generated by inversion breaking from substrates\cite{Gorbachev, Shimazaki, Sui} or strains\cite{Islam}. On the other hand, in heterostructures formed by TMD and other materials, interfacial Rashba SOCs can also emerge. For example, strong Rashba SOC has been reported recently in graphene/TMD hybrid structures\cite{Yang}. In the cases mentioned above, one can use moderate gating to tune the Fermi level in the range $\mu \sim 2|\beta^{\textrm{c}}_{\textrm{so}}|$ and study the unique $\sigma^{\textrm{V}}_{xy}-\mu$ curve due to SVHEs (Fig.\ref{FIG03}).

Third, the Berry phase in Eq.\ref{eq:03} for one K valley can also be generated in two-dimensional Rashba systems with large perpendicular magnetic field if orbital effects are ignored\cite{Culcer}. However, to generate a Zeeman splitting of a few meVs, an external magnetic field on the order of 100 Tesla is needed\cite{Ye1}. In such a strong magnetic field, the orbital effects cannot be ignored. Therefore, the family of TMDs with large Ising SOCs is very unique for demonstrating the novel SVHE.

Lastly, due to Ising SOC, valley Hall effects are generically accompanied by spin Hall effects in TMDs\cite{Xiao2}, which can also establish finite out-of-plane spin imbalance near the edges and contributes to polar Kerr effects. However, spin magnetic moments $\lesssim \mu_{\textrm{B}}$ ($\mu_{\textrm{B}}$: Bohr magneton) are generally small compared to the orbital valley magnetic moments $\sim 3-4$ $\mu_{\textrm{B}}$ in TMDs. Thus, polar Kerr effects are expected to be dominated by orbital magnetization.

\section*{\bf{Methods: Tight-binding Hamiltonian}}

In the Bloch basis of the following $d$-orbitals $\{\ket{d_{z^2,\uparrow}},\ket{{d_{xy,\uparrow}}},\ket{, d_{x^2-y^2,\uparrow}},\ket{d_{z^2,\downarrow}},\ket{ d_{xy, \downarrow}},\ket{d_{x^2-y^2,\downarrow}}\}$, the tight-binding Hamiltonian for gated/polar monolayer TMD is given by\cite{Liu}:
\begin{eqnarray}
H_{\text{TB}}\left(\bm{k}\right)&=&H_{\text{TNN}}\left(\bm{k}\right)\otimes\sigma_0 - \mu I_{6\times6}+\frac{1}{2}\lambda L_z\otimes\sigma_z \\\nonumber
&+& H_{\textrm{R}}(\bm{k}) + H^{\textrm{c}}_{\textrm{I}}(\bm{k}).
\end{eqnarray}

where
\begin{eqnarray}
H_{\text{TNN}}\left(\bm{k}\right)=\begin{pmatrix}
V_0 & V_1 & V_2 \\ 
V_1^{\ast} & V_{11} & V_{12} \\ 
V_2^{\ast} & V_{12}^{\ast} & V_{22}
\end{pmatrix}, & L_z=\begin{pmatrix}
0 & 0 & 0 \\ 
0 & 0 & -2i \\ 
0 & 2i & 0
\end{pmatrix}
\end{eqnarray}

refer to the spin-independent term and the atomic spin-orbit coupling term, respectively. $\mu$ is the chemical potential, and $I_{6\times6}$ is the $6\times6$ identity matrix. The Rashba SOC term is given by
\begin{eqnarray}
H_{\textrm{R}}(\bm{k}) =\text{diag}[2\alpha_0,2\alpha_2,2\alpha_2]\otimes(f_y(\bm{k})\sigma_x - f_x(\bm{k})\sigma_y). \end{eqnarray}

And the Ising SOC in the conduction band is given by
\begin{eqnarray}
H^{\textrm{c}}_{\textrm{I}}(\bm{k}) = \text{diag}[\beta(\bm{k}),0,0] \otimes \sigma_z.
\end{eqnarray}

Details of the matrix elements above can be found in Supplementary Note 3.

\section*{Data availability}
The data supporting the findings of this study are available within the paper, its Supplementary Information files, and from the corresponding author upon reasonable request.

\section*{Acknowledgement}
The authors thank Noah F. Yuan, C. Xiao and K. F. Mak for illuminating discussions. Y.T. was supported by Grant-in-Aid for Scientific Research on Innovative Areas ``Topological Materials Science" (KAKENHI Grants No.JP15H05851, No.JP15H05853, and No.JP15K21717), and for Challenging Exploratory Research (KAKENHI Grant No.JP15K13498). K.T.L acknowledges the support of Croucher Foundation, Dr. Tai-chin Lo Foundation and HKRGC through 16309718, 16307117, 16324216 and C6026-16W. 

\section*{Author contributions}
K.T.L conceived the ideas. B.T.Z. and K.T.L. prepared the manuscript. B.T.Z. and K.T. carried out the theoretical calculations. B.T.Z., K.T., Y.K. and Y.T. contributed to the analysis and interpretation of the results. All the authors discussed the results and commented on the manuscript.

\section*{Additional information}
Competing interests: The authors declare no competing interests.

\end{document}